# Subject integration with spreadsheets - Ignoring education is the greatest risk ever

Mária Csernoch, Ádám Gulácsi, Júlia Csernoch

University of Debrecen, Faculty of Informatics

csernoch.maria@inf.unideb.hu, gulacsi.adam@inf.unideb.hu, csernoch.julia@inf.unideb.hu

## ABSTRACT

Within the framework of Technological Pedagogical and Content Knowledge, subject integration is one possible solution for the introduction of meaningful digitalization and digitization in schools. This process incorporates that any school subject can be taught with digital support, informatics (computer) classes can be contextualized, and the gap between 'serious informatics' and 'digital literacy' can be minimized. The present paper details how three traditional Grade 3 tasks can be solved in spreadsheet, what skills, competencies, and computer science knowledge of both teachers and students can be developed. The solutions also reveal that analysing, understanding, planning, and discussing tasks is as important as the acting in the spreadsheets, which process plays a crucial role in the preparation of students for the future jobs.

## 1. INTRODUCTION

### 1.1. DidaTab

DidaTab (Didactique du Tableur, Didactics of Spreadsheet, Teaching and Learning Spreadsheets) (Blondel et al. 2008, Tort et al. 2009, Hérold & Montuori 2018) is one of the great projects in spreadsheet education

> "dedicated to the study of personal and classroom uses of spreadsheets in the French context, focusing on the processes of appropriation and uses by secondary school students."

The project revealed that spreadsheeting during secondary education (grade 6 to 12) is rather sparse for schoolwork (and even more seldom at home) and that student competencies are weak. It was also found that despite ICT is included in the prescribed curricula in many European countries, the ability to use professional software like spreadsheet is often based on self-studies and self-assessment (Brook et al. 2016, JRC 2022, Skov 2015). Consequently, the Dunning-Kruger Effect (Kruger & Dunning 1999, Dunning 2011) takes its toll, and these self-evaluation efforts draw a false image (Csernoch et al. 2015, 2021).

The DidaTab project concluded that

> "Use of spreadsheets is sparse. … These particular uses – that are not organised in time and occur at random (from a student point of view) – cannot give students the opportunity to build a sound knowledge of spreadsheets. … The episodic school activities involving spreadsheets are not sufficient to give students a clear view of what spreadsheets are and what can they be used for. These sparse uses are not sufficient for students to develop basic skills."

These conclusions were drawn after students attended 15 hours of spreadsheeting in Grade 7 Technology and some (the precise number is unknown) in Grades 8 and 9 Maths classes.

These hours compared to the requirements of the Hungarian National Base Curriculum would give opportunity and hope to build up some spreadsheet knowledge. In Hungary the first students meet spreadsheet is in Grade 8, and around 12 hours are assigned (suggested) to the subject. In Grades 9–10 and 11 further 12 + 12 hours are assigned to spreadsheeting. Nothing in Maths and other classes. Without going into details, during these 12 + 12 + 12 hours a wide range of spreadsheet features should be taught, including 66 functions. The number of functions alone, 2 per classes, indicates that the cognitive load is extremely high, the curricula do not allow space for learning (Csernoch & Biró 2017).

We argue, based on the results of DidaTab and the introduction of spreadsheeting in Hungarian schools, that learning spreadsheeting should start as soon as students can read. One of the advantages of this approach is to reduce cognitive load by distributing knowledge items while another is to switch the focus of the teaching-learning process from tools to humans. Methods based on these values are motivating, reduce cognitive load, and consequently allow us to build up fundamental Computer Science (CS) knowledge.

We also state that the illusion of teaching – introducing as many features (tools) of a software as the pages of coursebooks allow – is one of the greatest risks. With this approach, on one hand the essence of teaching informatics, the transfer of fundamental CS knowledge is lost. On the other hand, all participants of the learning-teaching process – students, teachers, parents, education leaders, etc. – are misled and a false image of knowledge is built up, on which future jobs cannot rely.

### 1.2. Future Jobs

According to the Future of Jobs Report 2023, the top core skills required by workers today are analytical thinking, creative thinking, self-efficacy skills – resilience, flexibility, and agility –, motivation and self-awareness, curiosity and lifelong learning, technological literacy, dependability, and attention to detail (The Future of Jobs Report 2023). The question is how schools and trainings can prepare future employees to fulfil the requirements of the jobs in the shadow of digitalization and digitization. However, schools do not seem to be ready for this challenge. Primarily, education tends to separate computer studies (informatics) from digital literacy (Caspersen et al. 2022, Vahrenhold 2017), not seeing that computational thinking skills should be developed in union (Soloway 1993) to learn computer science. DigCompEdu (Redecker 2017) goes even further by claiming that data is not considered as digital resources in education because they should be analysed. With this approach, development of the analytical skill, in the first position is ruled out.

> "Digital resources: The term usually refers to any content published in computer-readable format. For the purposes of DigCompEdu, a distinction is made between digital resources and data. Digital resources in this respect comprise any kind of digital content that is immediately understandable to a human user, whereas data need to be analysed, treated and/or interpreted to be of use for educators." (Redecker 2017 pp:90)

Unfortunately, this idea is completely supported by the tool-centred, digital push education systems (Modig & Åhlström 2018), where hardware and software tools are in the focus, instead of the development of problem-solving skills, computational thinking skills (Wing 2006), along with the above-mentioned core skills of the students (The Future of Jobs Report 2023). According to the results of our research project, subject integration might be one possible solution for the fulfilment of the future jobs requirements. On the one hand, subject integration is open to use digital solutions where traditional, paper-based solutions do not motivate students anymore and where digitization might develop skills and knowledge which were never dreamed of (Vegas et al. 2021, Wolfram 2020). On the other hand, subject integration can provide content to computer education, consequently, misconceptions that the office software suit should be banished from education can be revisited (Gove 2012, Denning 2013).

> "Instead of children bored out of their minds being taught how to use Word and Excel by bored teachers…" (Gove 2012).

> "From early analyses, we could see that students were losing interest in computing in high schools, half of which had no computer course at all, and many of the others relegated their one computer course to literacy in keyboarding and word processing." (Denning 2013)

Finally, but not least, subject integration within computer studies would be able to close the gap between 'serious informatics' and 'digital literacy' (Soloway 1993, Guzdial & Soloway 2002).

The Technological Pedagogical And Content Knowledge (TPACK) (Figure 1) provides the theoretical framework for digital subject integration (Mishra & Koehler 2006, Angeli & Valanides 2015).

> "At the heart of the TPACK framework, is the complex interplay of three primary forms of knowledge: Content (CK), Pedagogy (PK), and Technology (TK). The TPACK approach goes beyond seeing these three knowledge bases in isolation. The TPACK framework goes further by emphasizing the kinds of knowledge that lie at the intersections between three primary forms: Pedagogical Content Knowledge (PCK), Technological Content Knowledge (TCK), Technological Pedagogical Knowledge (TPK), and Technological Pedagogical Content Knowledge (TPACK). Effective technology integration for pedagogy around specific subject matter requires developing sensitivity to the dynamic, transactional relationship between these components of knowledge situated in unique contexts." (Koehler 2012)

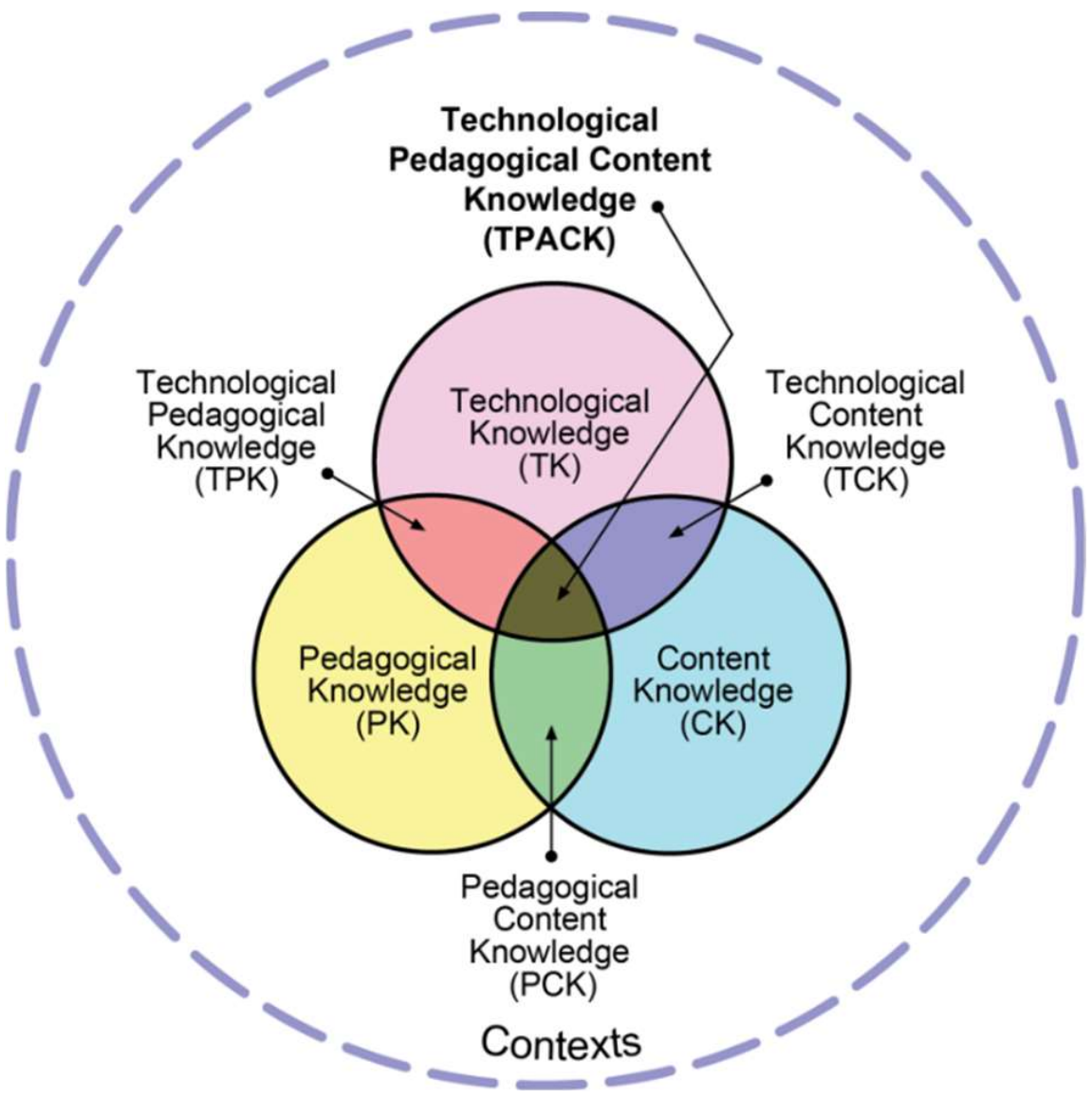

Figure 1. The TPACK framework[1].

---

[1] http://matt-koehler.com/tpack2/wp-content/uploads/2013/08/TPACK-new-768x768.png

With this background there are various solutions for subject integration, however in the present paper we focus on how spreadsheet can be used effectively and efficiently in the development of primary school students' core skills.

### 1.3. Spreadsheet subject integration

In accordance with Math(s) Fix (Wolfram 2020) and TPACK, digital subject integration can help to develop skills, competencies, and knowledge that lie in the intersection of technology, pedagogy, and content. It is remarkable that in school curricula, spreadsheeting is considered a special tool which is not allowed to be introduced sooner than Grades 7 or 8 or something that is not suitable for students and should be banished from schools (Gove 2012). We argue that it is not so, the programming approach of spreadsheeting can be introduced as soon as students can read (Sestoft 2011, Csernoch 2014), and suitable, meaningful tasks can be motivating and interesting for small children. We must also call attention to the problem-solving approach applied to these activities. The tool-centred approaches allow students and end-users to jump headfirst into computers and applications, leaving out the crucial steps of problem solving (Polya 1945, Smalley 2018). In the present work, we emphasize the importance of data analysis, planning, the details of acting, and discussion.

Spreadsheeting and TPACK primarily considered as methods to develop maths skills and knowledge (Niess 2005, Agyei & Keengwe 2012, Agyei & Voogt 2012), "because it has potential for supporting students' higher-order thinking in mathematics" (Agyei & Keengwe 2012). However, spreadsheets can be connected to other school subjects and sciences, which we plan to demonstrate. However, we must call attention to the trend where spreadsheets are primarily meant to students in higher grades, hardly mentioned in primary education (References 2023).

In the following, the details of three conversions are presented where the tasks are selected from Grade 3 course books (Fülöp & Szilágyi 2022 pp:29 and Tóthné & Vitéz 2022 p:26 and pp:46). It is analysed what skills, competencies, and knowledge can be developed focusing on the spreadsheet solutions. It is discussed how the selected tasks can be built upon each other, and how teachers can or should prepare the tasks in accordance with the student's background knowledge and the target conditions of the tasks and classes.

## 2. SPREADSHEET SOLUTIONS TO PAPER-BASED TASKS

For the present paper three tasks are selected to show how programming-oriented spreadsheeting can be introduced in Grade 3 classes (Sestoft 2011, Csernoch 2014). One task arrives from a grammar book (Fülöp & Szilágyi 2022 pp:29) and two from a science book (Tóthné & Vitéz 2022 p:26 and pp:46). It is the decisions of the schools and the teachers in which classes these problems are solved whether in grammar/sciences or informatics classes. It always depends on the teachers' digital skills (technological knowledge), content knowledge, and pedagogical, didactical readiness. According to TPACK, to be effective in digital problem-solving, all three components should be present in these classes.

### 2.1. Reading comprehension exercise – morning routine

The original task requires students to recognize the correct order of morning routine tasks by reading the sentences and then writing a number to each task (Figure 2).

#### 2.1.1. Paper solution

The paper-based solution requires the reading of the text and then a paper and a pencil, optionally an eraser, to write the numbers.

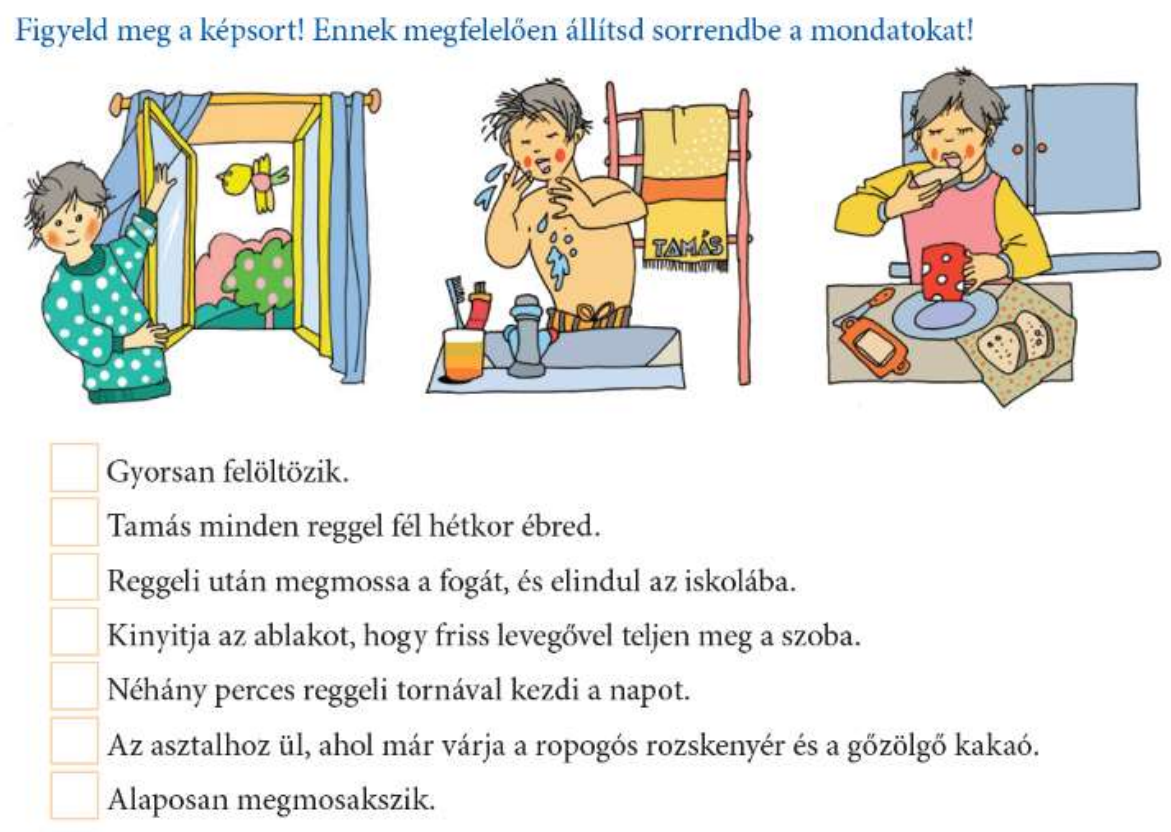

Look at the sequence of pictures! Put the sentences in order accordingly.

- ☐ He dresses quickly.
- ☐ Thomas wakes up every morning at 6:30.
- ☐ After breakfast, he brushes his teeth and goes to school.
- ☐ He opens the window to let fresh air into the room.
- ☐ He starts the day with a few minutes of morning exercise.
- ☐ He sits at the table, where crusty rye bread and steaming cocoa await him.
- ☐ He washes himself thoroughly.

Figure 2. The original Hungarian paper-based (left) reading comprehension task and its translation (right).

### 2.1.2. Designing a spreadsheet solution

The digital solution requires also the reading and understanding of the text, however solving the problem in Excel calls for some planning. In the planning phase students should recognize that a new Excel workbook would be created, so the name of the folder, the file, and the worksheet should be discussed and set. After that students would decide on the number of fields, the cells where the text goes, and how the text can be entered into those cells. A hand-drawn tagged plan would serve our purposes the best.

### 2.1.3. A spreadsheet solution

One possible solution is presented in Figure 3. In the first step the field names should be typed. This step requires the selection of the cell without – positioning the cursor into the cell – and moving from one cell to the other (upper left image). In the second step the unformatted text (Text only) is copied from the course book into Cell B2 of the worksheet, which is a meaningful (contextualized) task to introduce and practice copying and its algorithm. At this phase, a short data analyses should follow, which allows students to recognize that each paragraph (sentence) is copied to a different cell, consequently as many records are created as the number of sentences in the exercise. Some students can also figure out that Column B is not wide enough, so the changing of column width to fit the content (column separator mouse pointer and double click) can also be introduced. The panes can also be frozen, which is an additional useful piece of information.

When the table is set up properly, students can type their numbers into the corresponding cells. Here, they can also practice that for typing, only the cell should be selected – the cursor is not needed – and moving along the vertical array is faster and safer with the up and down keyboard arrows than with the mouse. Result is presented in Figure 3 (upper right image). However, most importantly, students' attention must call to the alignment of both number and text. They should recognize that text is aligned left and number right automatically. The role of the teacher is to emphasize that we do not change the alignment, because they help us to recognize the data type assigned to the entered data.

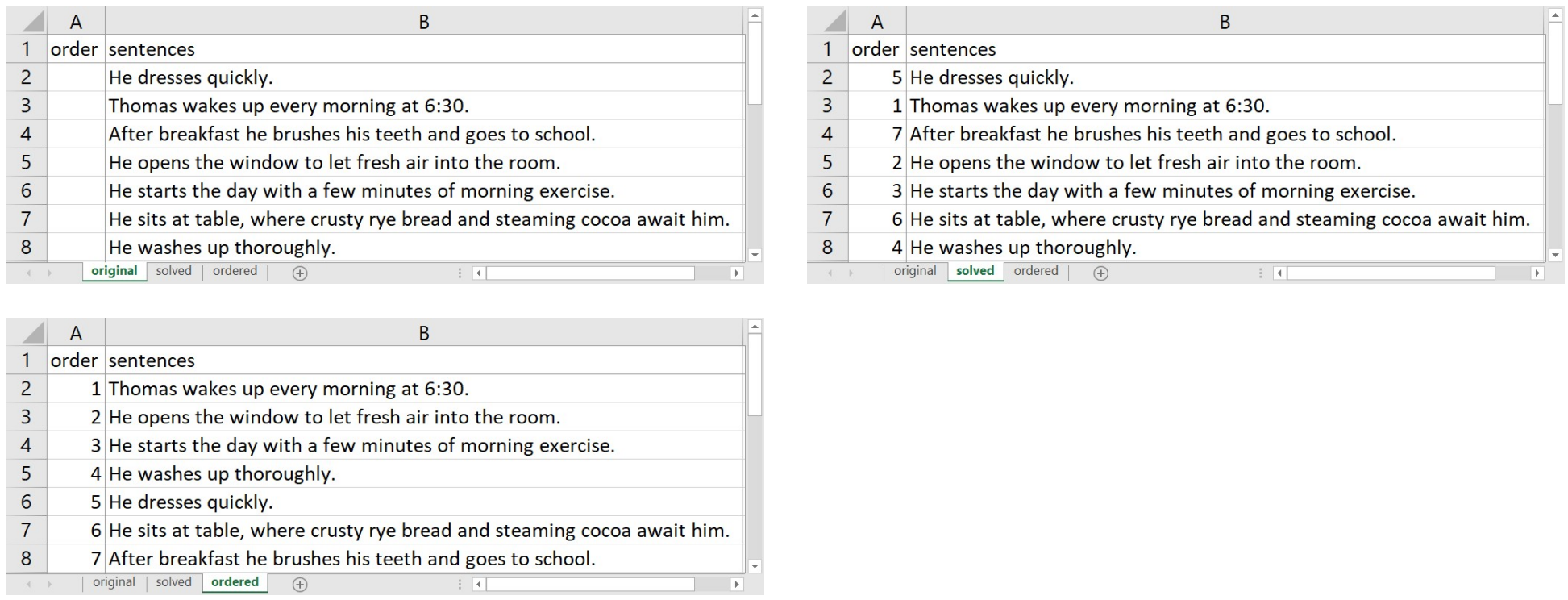

| | A | B |
|---|---|---|
| 1 | order | sentences |
| 2 | | He dresses quickly. |
| 3 | | Thomas wakes up every morning at 6:30. |
| 4 | | After breakfast he brushes his teeth and goes to school. |
| 5 | | He opens the window to let fresh air into the room. |
| 6 | | He starts the day with a few minutes of morning exercise. |
| 7 | | He sits at table, where crusty rye bread and steaming cocoa await him. |
| 8 | | He washes up thoroughly. |

original | solved | ordered

| | A | B |
|---|---|---|
| 1 | order | sentences |
| 2 | 5 | He dresses quickly. |
| 3 | 1 | Thomas wakes up every morning at 6:30. |
| 4 | 7 | After breakfast he brushes his teeth and goes to school. |
| 5 | 2 | He opens the window to let fresh air into the room. |
| 6 | 3 | He starts the day with a few minutes of morning exercise. |
| 7 | 6 | He sits at table, where crusty rye bread and steaming cocoa await him. |
| 8 | 4 | He washes up thoroughly. |

original | solved | ordered

| | A | B |
|---|---|---|
| 1 | order | sentences |
| 2 | 1 | Thomas wakes up every morning at 6:30. |
| 3 | 2 | He opens the window to let fresh air into the room. |
| 4 | 3 | He starts the day with a few minutes of morning exercise. |
| 5 | 4 | He washes up thoroughly. |
| 6 | 5 | He dresses quickly. |
| 7 | 6 | He sits at table, where crusty rye bread and steaming cocoa await him. |
| 8 | 7 | After breakfast he brushes his teeth and goes to school. |

original | solved | ordered

Figure 3. The three phases of the solution of the reading comprehension task in a spreadsheet workbook.

### 2.1.4. Let's do the magic!

After typing the numbers, spreadsheet can rearrange the sentences based on the ranking. At this step students can learn that the selected cell (again, without the cursor) should be in the range of the numbers – there is no need for the selection of the whole matrix – and the *Sort A to Z* command will do the magic. After the command is evaluated, the numbers are in ascending order and sentences appear in the order of the morning tasks (Figure 3, lower left image).

## 2.2. Estimation task – school objects

Another classic example of subject integration is estimation (Figure 4). The task is found in the Grade 3 Sciences course book (Tóthné & Vitéz 2022 p:26), but mathematics plays a just as important role as science. First, students are asked to estimate the length of the presented four objects expressed in cm, then they should measure them. The answers should be written on the dotted lines. This exercise can be turned into an Excel solution.

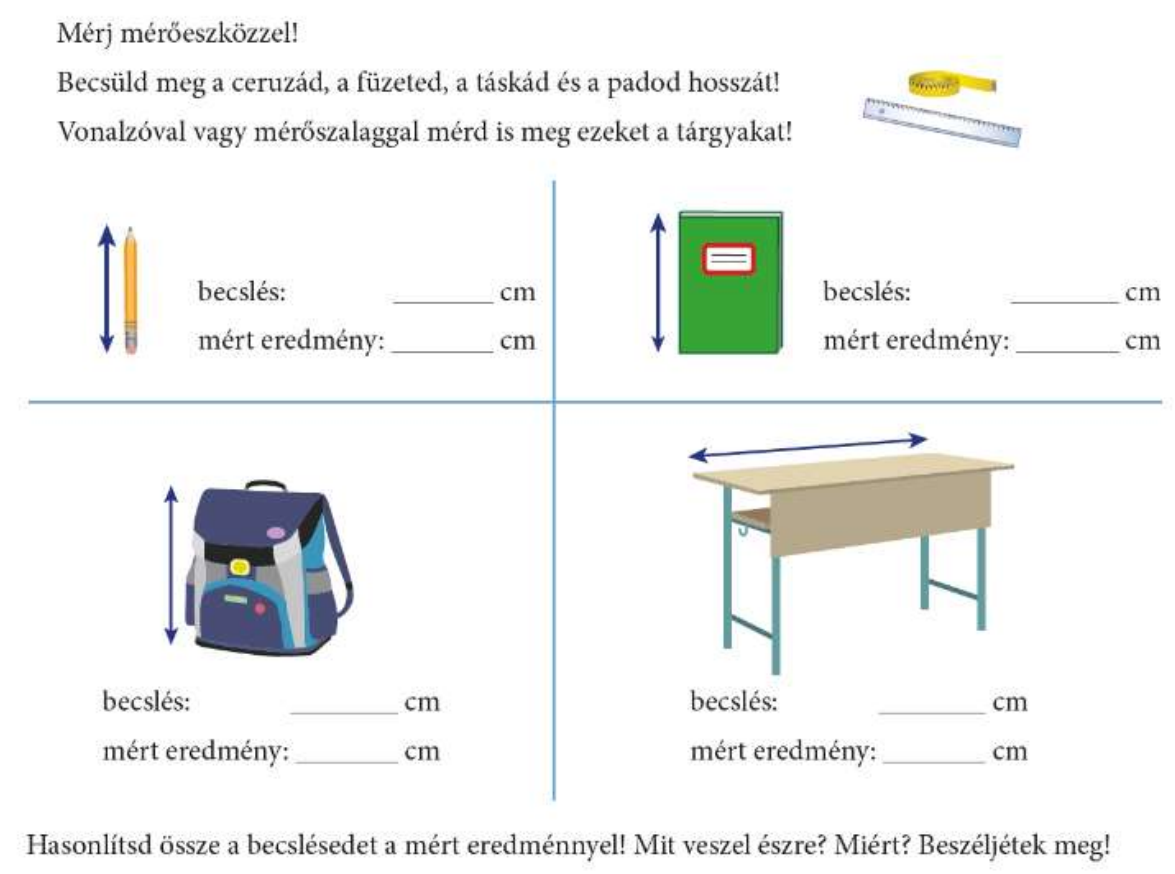

Measure with a meter!

Estimate the length of your pencil, notebook, bag, and desk.

Use a ruler or tape measure to measure these items too!

estimate: cm

measured result: cm

Compare your estimate with the measured result! What do you notice? Why? Discuss!

Figure 4. The original Hungarian paper-based (left) estimation task and its translation (right).

### 2.2.1. Designing the spreadsheet table

Similar to the previous task, this one should be planned on a piece of paper in advance to any act in a spreadsheet (the saving of the file is not detailed any more). Based on the original task, we need

- three fields to enter the name of the objects, the estimated and the measured values,
- four records, and

- one row at the top of the table for the field names.

We should keep in mind that the worksheet should also be named.

### 2.2.2. Creating the spreadsheet table

The typing load is reasonable, even for Grade 3 students (Figure 5, upper left image). The alignment of the strings and numbers plays a crucial role, as it is mentioned in the previous task. However, in this case, students want to type the cm with the number. Teachers must find ways to convince students that typing the cm changes the type of the data, and the formatting of the numbers is the correct solution (Figure 5, upper right image), not their rearrangement (e.g., a SUM() function on the teacher's computer might be enough).

Once the numbers are correctly formatted, the following step is the calculation of the differences. To display the results, we need a new field. This new field should be added to the plan. This is one of the first formulas of the students, where a substruction should be carried out (Figure 5, lower left image). If they are lucky, they have Office 365 which handles dynamic arrays. If not, it is up to the teacher's decision whether they copy the first formula downwards or use array formula.

| | A | B | C | D | E |
|---|---|---|---|---|---|
| 1 | object | estimate | measure | | |
| 2 | pencil | 15 | 13 | | |
| 3 | notebook | 20 | 19 | | |
| 4 | bag | 35 | 42 | | |
| 5 | desk | 100 | 125 | | |

original | unit | difference1 | difference2

| | A | B | C | D | E |
|---|---|---|---|---|---|
| 1 | object | estimate | measure | | |
| 2 | pencil | 15 cm | 13 cm | | |
| 3 | notebook | 20 cm | 19 cm | | |
| 4 | bag | 35 cm | 42 cm | | |
| 5 | desk | 100 cm | 125 cm | | |

original | unit | difference1 | difference2

| | A | B | C | D | E |
|---|---|---|---|---|---|
| 1 | object | estimate | measure | difference | |
| 2 | pencil | 15 cm | 13 cm | 2 cm | |
| 3 | notebook | 20 cm | 19 cm | 1 cm | |
| 4 | bag | 35 cm | 42 cm | -7 cm | |
| 5 | desk | 100 cm | 125 cm | -25 cm | |

original | unit | difference1 | difference2

| | A | B | C | D | E |
|---|---|---|---|---|---|
| 1 | object | estimate | measure | difference1 | difference2 |
| 2 | pencil | 15 cm | 13 cm | 2 cm | 2 cm |
| 3 | notebook | 20 cm | 19 cm | 1 cm | 1 cm |
| 4 | bag | 35 cm | 42 cm | -7 cm | 7 cm |
| 5 | desk | 100 cm | 125 cm | -25 cm | 25 cm |

original | unit | difference1 | difference2

Figure 5. The four phases of the solution of the estimation task in a spreadsheet workbook.

### 2.2.3. Discussion

The task can be considered completed when all the differences are calculated, but teachers can think greater and adventure into to the field of absolute value. Grade 3 students in normal class do not study the absolute value, but with the ABS() function, the magic would work again (Wolfram 2020). To calculate the absolute values, students need one more field, and they can call their first function (Figure 5, lower right image).

## 2.3. Fishing problem – diagram

Reading and drawing diagrams require skills and competences which most students struggle with. One of the reasons might be the complexity of the diagrams generated from the several subjects involved in. For the next example a complex problem is selected from the Grade 3 Sciences book (Tóthné & Vitéz 2022 pp:46). The accompanying diagram is presented in Figure 6, the fishing rules (conditions) in Figure 14, and additional questions in Figure 15 (Figure 16), where the answers should be formulated based on the diagram and the conditions.

### 2.3.1. Reading a diagram

The first task of the fishing problem is to read the diagram and write the lengths of the caught fish[2]. In the original task the empty slots right to the diagram are offered to write the numbers.

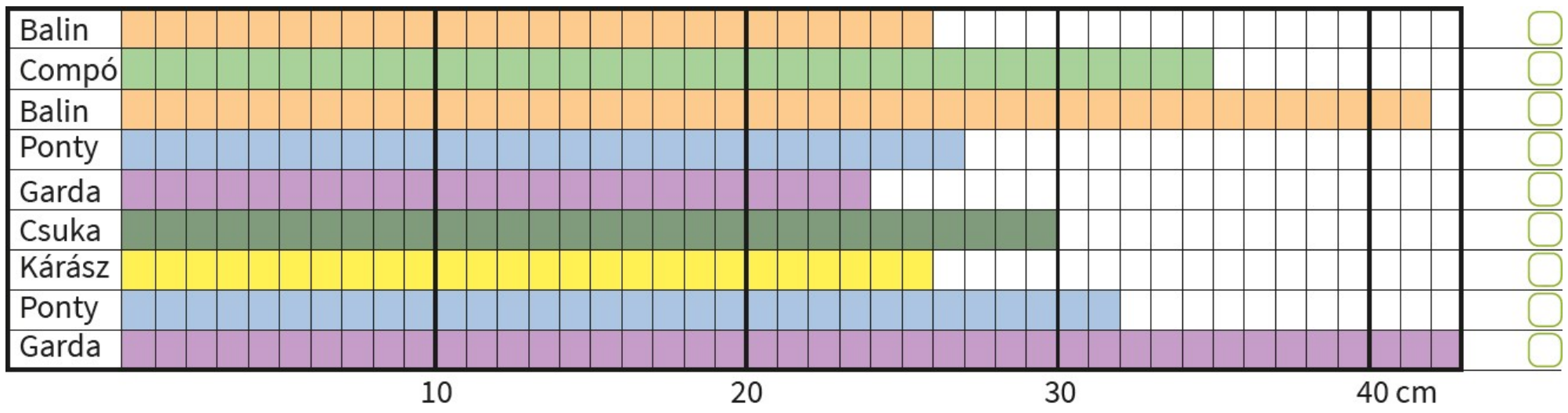

Figure 6. The original Hungarian paper-based diagram of the fish problem.

### 2.3.2. Planning the table and entering data

If we turn to Excel to record the result, we need two fields, the name of the fish (fish) and the size (size) read from the original diagram (Figure 6) in the course book. In the following, the format of the numbers should be set to cm (Figure 7).

| | A | B |
|---|---|---|
| 1 | fish | size |
| 2 | balin | 26 cm |
| 3 | compó | 35 cm |
| 4 | balin | 42 cm |
| 5 | ponty | 27 cm |
| 6 | garda | 24 cm |
| 7 | csuka | 30 cm |
| 8 | kárász | 26 cm |
| 9 | ponty | 32 cm |
| 10 | garda | 43 cm |

Figure 7. The caught fish and their size are recorded, and the numbers are formatted as cm.

### 2.3.3. Skills and competences

In our course plan, the fish problem is the third one where spreadsheet can be used (entering data: copying or typing; manipulating data: ordering, using an operator in a formula, calling function). At this stage, students already have some experience with fundamental spreadsheet concepts and practices.

Based on the data presented in the course book and the students' previous exercises creating and formatting a diagram would be a challenging but motivating task.

### 2.3.4. Planning a diagram

One might argue that Grade 3 students cannot create a diagram of such complexity. However, in accordance with Wolfram, computers are meant to replace boring, repetitive tasks and introduce concepts are higher level (Wolfram 2020) and in accordance with kaizen, small steps would cause changes never seen before.

---

[2] fish, chub, ide, carp, bleak, pike, roach (translated by ChatGPT)

We must keep in mind that Grade 3 students have neither design nor diagram practices. This includes that in advance to creating the diagram, the traditional planning phase should be reconsidered. In this case, we can start with the most basic diagram, analyse the result, do a comparison to the original diagram, and decide on the next step. This cycle can be repeated until we reach the appearance closest to the original diagram. This method allows us to discuss sets associated with diagrams (Domain, Range), the advantages of info graphics, the diagram related tools of spreadsheets, colour coding, etc.

### 2.3.5. Creating and formatting a diagram

The first step is to create the simplest 2-D Bar diagram based on the size of the fish (Figure 8, left). The analysis of the diagram reveals that the values (sizes) (Range) are assigned to the horizontal axis, while the vertical axis has numbers instead of the names of the fish (Domain). In the next step we set up the correct labels on the vertical axis by adding the fish field to the diagram (Figure 8, right).

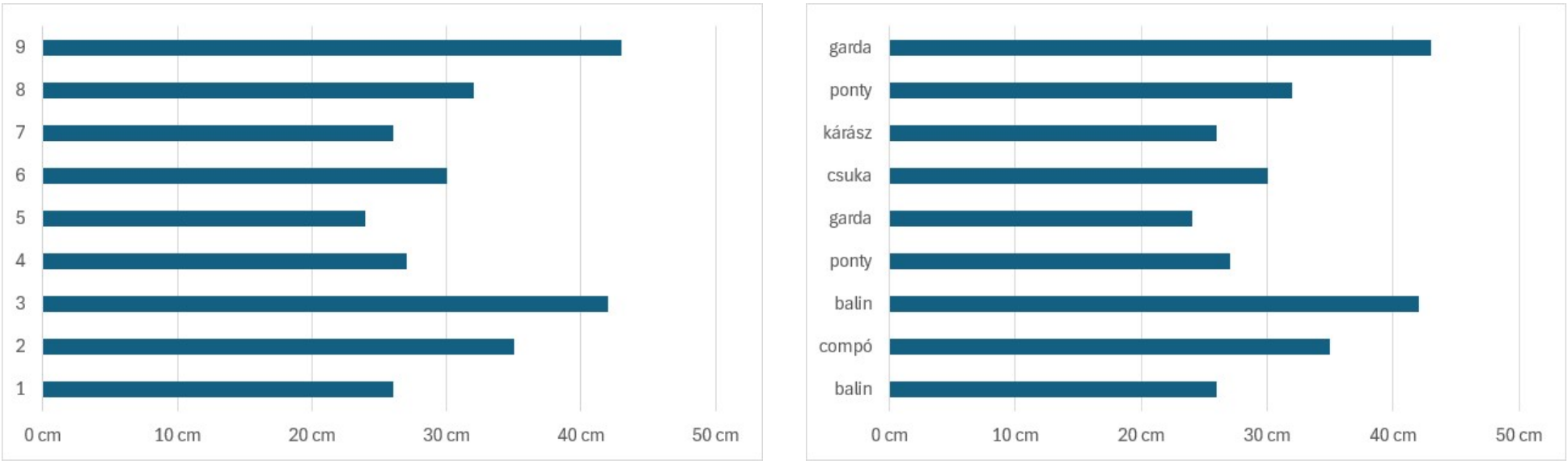

Figure 8. Inserting a 2-D Bar diagram (left) and adding the Category Axis Labels in Select Data Source (right).

The following analysis reveals that we should change the order of the fish on the vertical axis (Figure 9, left) and the maximum value of the horizontal axis to 45 in accordance with the original diagram (Figure 9, right).

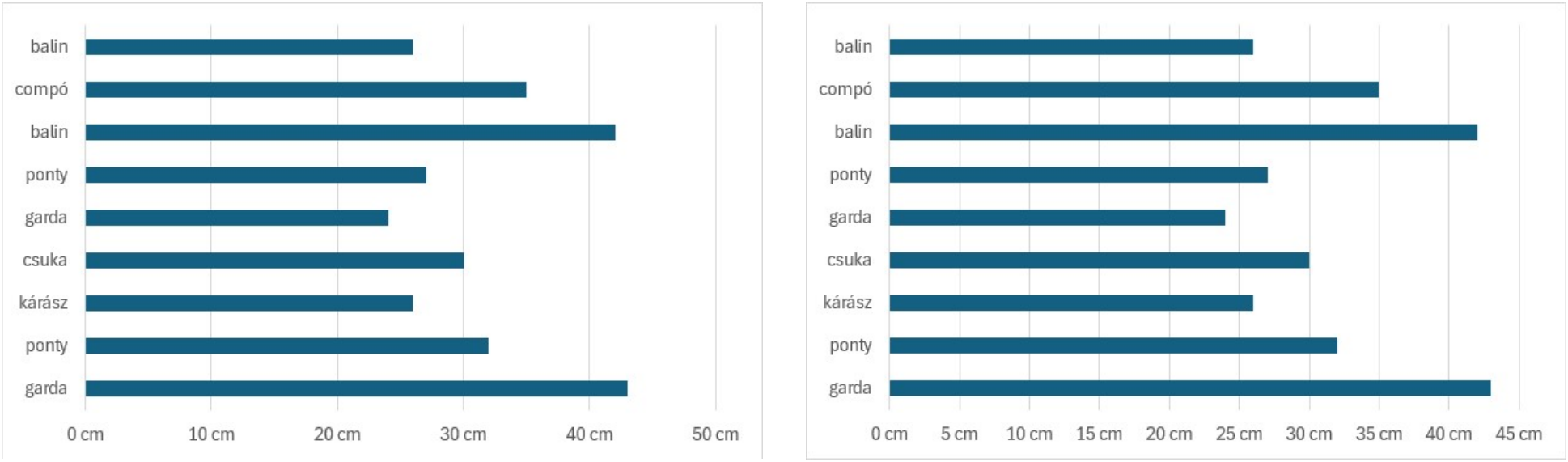

Figure 9. Changing the order of the fish (left) and the maximum value of the horizontal axis (right).

The diagram has bars according to the size of fish, labels on both invisible axes, and vertical grids in every 10 cm. However, it does not have axes and horizontal grids, and the vertical grids are not frequent enough. In the next step the axes (Figure 11, left) and then the gridlines can be displayed (Figure 11, right).

| fish | R | G | B |
|---|---|---|---|
| balin | 252[3] | 203 | 141 |
| compó | 169 | 210 | 154 |
| ponty | 171 | 197 | 226 |
| garda | 198 | 156 | 200 |
| csuka | 128 | 154 | 124 |
| kárász | 255 | 241 | 85 |

Figure 10. The RGB codes of the diagram.

After the axes and gridlines appear on the diagram, the colours of the bars should be changed. This is a two-step process. First the RGB codes of the colours should be collected, which can be completed by either the teacher in advance to the class or the students during the class. It is always the teacher's decision, in the function of the target condition of the task (Modig & Åhlström 2018). The second step is to change the fill colour of the bars in the spreadsheet. To collect the RGB codes of the colours, the picture of the original diagram should be opened in a graphical software and the colour of each bar should be required by a colour picker tool (Figure 10). To change the fill colour of the bars, the collected RGB code should be entered in either the *Custom Color* dialog box or selected from the *Resent Colors* collection (Figure 11, right).

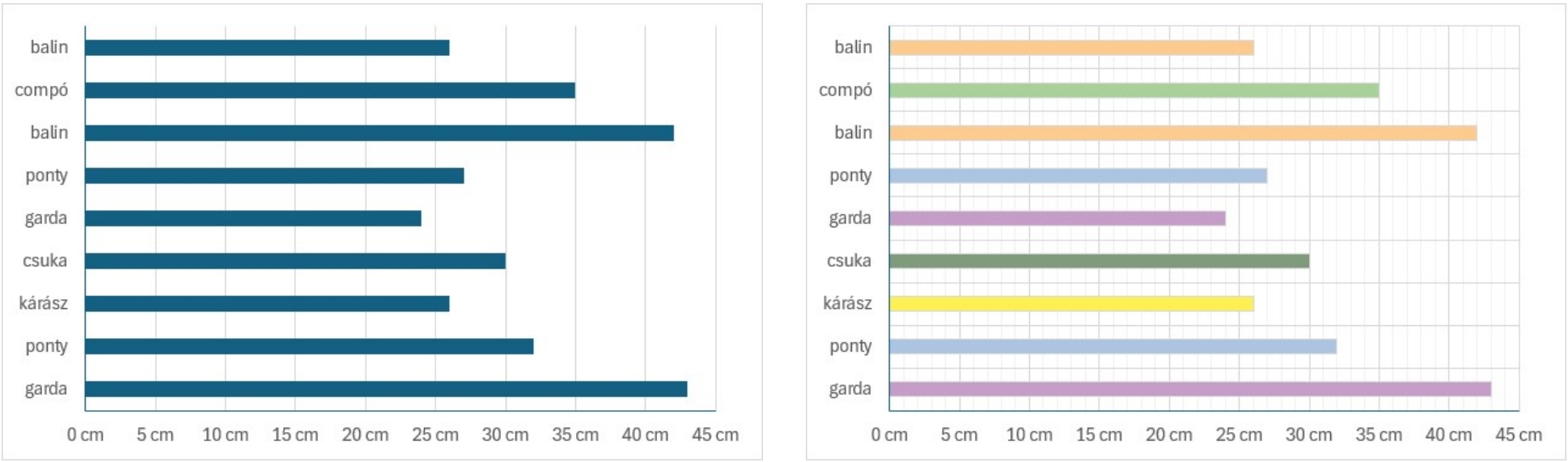

Figure 11. The axes (left) and gridlines (right) are displayed, and the bars are coloured (right).

In the following step, the gap between the bars should be minimized. Excel allows us to change the Gap With to 0% (Figure 12, left). The diagram is ready, but we cannot see the grids, and it is difficult to tell the right size of the fish without these grids. However, the only tool which we found to display the grids is the transparency of the bars (Figure 12, right).

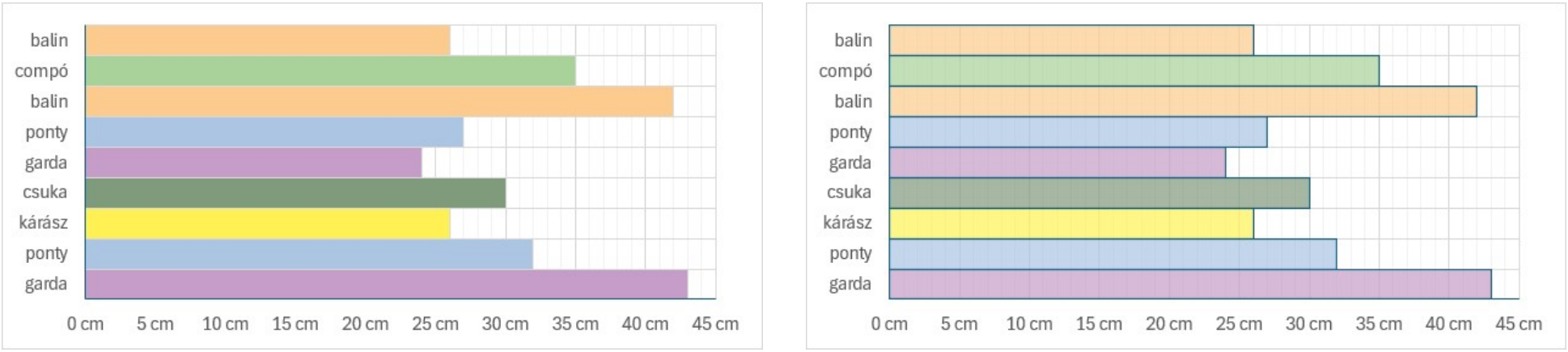

Figure 12. The gap between the bars is set 0% and the transparency of the bars to 30%.

### 2.3.6. Discussion

As we realized, the grids cannot be placed on the top of the bars. However, the comparison of the original and the Excel diagram reveals that there are further differences. The alignment of the fishes on the vertical axis are different, there is no room for the empty slots to enter the numbers, and the squares on the right side cannot be added to the diagram. A thorough analysis of the formatting of the Excel diagram and the appearance of the original leads us to the conclusion that the original diagram is fake. It is not a diagram, but a table, which can be created in Excel, Word, PowerPoint, considering the Office suit. In the following, the Excel version of the table is presented.

[3] RGBs are collected with the Color Picker Tool of GIMP

## 2.4. Fishing task – fake diagram

2.4.1. Formatting a table

From now on, the diagram task is handled as a formatting exercise, which requires the following settings.

- creating squares (row height = column width)
- colouring the right number of squares with the RGB codes presented in Figure 10
- setting the alignment of the fish (horizontal: left, vertical: middle)
- adding borders (according to the sample)
- inserting rounded green squares – RGB(166,193,88) – into cell, alignment (right)
- copying the cells with the squares
- adding labels to the horizontal axes, merging cells

The result is presented in Figure 13.

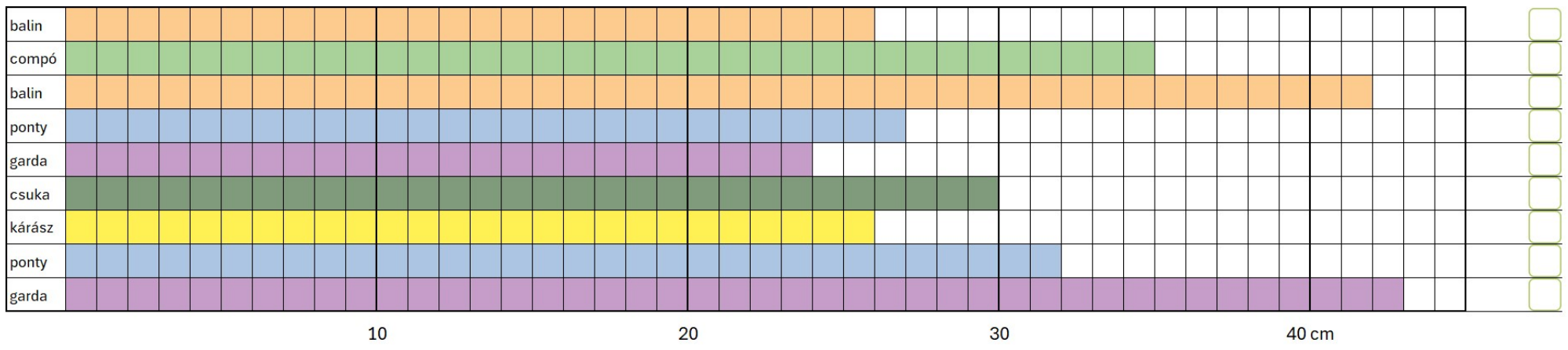

Figure 13. The fake diagram = formatted table created in a spreadsheet.

### 2.4.2. Discussion

This formatting exercise fits Grade 3 students background knowledge and serves subject integration within the Office suite. These formatting steps can be applied in similar tasks and a good starting point for handling tables. The different solutions in the different programs might reveal similarities and differences which can be discussed in the course of the learning-teaching process.

### 2.5. Fishing task – further conditions and questions

Based on the conditions (Figure 14) and the diagram (Figure 6), further questions are asked in the fishing problem (Figure 15 and Figure 16). The spreadsheet solution of this extension is way beyond the knowledge and skills of Grade 3 students but suitable and challenging for higher grades.

### 2.5.1. Conditions

There are three conditions on which fish can or cannot be taken (must be thrown back) (Figure 14). There is limit on the size and the number of fish for each fish, and on the total number of fish.

| Megnevezés | Legkisebb kifogható méret | Naponta kifogható mennyiség | Naponta összesen kifogható mennyiség |
|---|---|---|---|
| Balin | 40 cm | 3 db | 5 db<br>A kifogott halat korábban kifogott halra kicserélni tilos! |
| Compó | 25 cm | 3 db | |
| Csuka | 40 cm | 3 db | |
| Garda | 20 cm | – | |
| Kárász | Méretkorlátozás nincs. | – | |
| Ponty | 30 cm | 3 db | |

| Name | Smallest size that can be caught | Daily catch quantity | Total per day catches |
|---|---|---|---|
| balin | 40 cm | 3 pieces | 5 pieces<br>It is prohibited to exchange a caught fish for a previously caught fish. |
| compó | 25 cm | 3 pieces | |
| csuka | 40 cm | 3 pieces | |
| garda | 20 cm | No limit. | |
| kárász | No size limit. | No limit. | |
| ponty | 30 cm | 3 pieces | |

Figure 14. The original Hungarian conditions of the fishing problem (left) and its translation (right).

### 2.5.2. Questions

The questions are presented in Figure 15 (Figure 16). To answer these questions high level programming and/or spreadsheet skills are required. The complexity of the spreadsheet solution depends on the number of constant values and variables which users are allowed to use. Furthermore, multiple conditional formatting rules can be used to highlight the answers and make the spreadsheet dynamic.

*a)* Tanulmányozd a grafikont! Melyik hal mekkora volt? Írd a sorok után!

*b)* Ha hazavitték, írj H betűt, ha visszadobták, írj V betűt a sor végére!

*c)* Igaz vagy hamis az állítás?

Több halat vittek haza, mint amennyit visszadobtak. ☐

Az azonos méretű halak egyforma sorsra jutottak. ☐

*d)* Adj választ a kérdésekre!

- Melyik halból nem vihettek haza?
- Melyik hal volt a legnagyobb zsákmány?
- Mekkora volt a legkisebb hal, amit hazavihettek?
- Milyen hal volt a legnagyobb, amit visszadobtak?

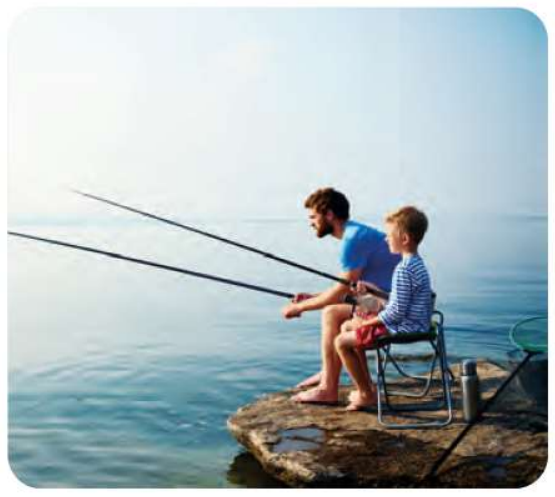

Figure 15. Further questions of the fishing problem.

a) Study the graph! What is the size of caught fish? Write after the lines!

b) If it was taken home, write an H, if it was thrown back, write a V at the end of the line!

c) Is it true or false?

They took home more fish than they threw back. ☐

Fish of the same size met the same fate. ☐

d) Answer the question!

- Which fish they could not take home?
- Which fish was the biggest prey?
- What was the size of the smallest fish they could take home?
- What was the biggest fish they threw back?

Figure 16. Further questions of the fishing problem in English.

### 2.5.3. Discussion

We must call attention to the conditions. It is not clear what the dashes mean in the third column. We found at least two explanations: (1) there is no limit on the number of caught fish or (2) that fish cannot be caught in this competition.

The number of constant values and variables allowed in the spreadsheet make the solutions simpler but less flexible, and the other way around. The task can be further diversified whether substitute cells and ranges are allowed or not.

It is also not clear in Question 2 in d) what prey means. All the caught fish should be taken into consideration or only those which were taken home.

Since the largest number of the caught fish is 3, the daily limit on the number of fishes does not count.

One possible solution is presented in Figure 17.

| | A | B | C | D | E | F | G | H | I | J | K | L | M | N | O | P |
|---|---|---|---|---|---|---|---|---|---|---|---|---|---|---|---|---|
| 1 | fishes | smallest | pieces per_day | FC3 | SC3 | FC4 | SC4 | taken | left | no_catch | largest1 | largest2 | smallest taken | largest left | smallest taken | largest left |
| 2 | balin | 40 cm | 3 pc | L | T | 26 | 42 | 1 | 1 | | 42 | balin | 42 | 26 | | |
| 3 | compó | 25 cm | 3 pc | T | | 35 | | 1 | 0 | | 35 | | 35 | 0 | | |
| 4 | csuka | 40 cm | 3 pc | L | | 30 | | 0 | 1 | csuka | 0 | | 0 | 30 | | |
| 5 | garda | 20 cm | – | T | L | 24 | 43 | 1 | 1 | | 24 | | 24 | 43 | garda | garda |
| 6 | kárász | – | – | T | | 26 | | 1 | 0 | | 26 | | 26 | 0 | | |
| 7 | ponty | 30 cm | 3 pc | L | T | 27 | 32 | 1 | 1 | | 32 | | 32 | 27 | | |
| 8 | | | | | | | | 5 | 4 | | | | 24 | 43 | | |

size | size_diagram | size_diagram2 | size_diagram3 | size_diagram3_fit | taken1 | taken2 | conditions1 | **conditions2** | questions

Figure 17. Answering the questions with a spreadsheet solution.

## 3. CONCLUSIONS

In the present paper the digitization of three traditional paper-based tasks, arriving from Grade 3 course books, is detailed. On the one hand, the original tasks are converted into spreadsheet solutions proving that Grade 3 students can handle spreadsheets if the requirements of the tasks are in accordance with their skills and knowledge. On the other hand, along with the solutions of the tasks it is also discussed what skills, competencies, and knowledge of students and teachers can be developed, including teachers' methodological, pedagogical, and didactical skills (Stigler & Hiebert 2009, Hattie 2012). It is also mentioned how these tasks can be further developed serving higher grades and how teachers can prepare them in accordance with the target conditions of the tasks and classes.

Cognitive load (Sweller 2011) plays a crucial role in the development of students' skills and knowledge. In the tool-centred digital education systems (Modig& Åhlström 2018), tons of tools are presented in the hope of building up huge knowledge inventories. However, our long-term memory cannot work as an inventory, we forget, especially decontextualized data are lost. The presented tasks are selected to avoid these losses and only those tools are introduced which are necessary to solve the problems, in accordance with a pull digital education concept. The aims of these digitization procedures are to build up long-lasting knowledge by providing opportunities to analyse data, to be creative, to practice and repeat actions in spreadsheet, to pay attention to details, to build up connections between pieces of information arriving from various sources, to motivate students, to give meaning of informatics (computers: hardware and software tools). These skills play a crucial role in the development of students and schools since they are in complete accordance with the requirements of the future jobs.